\documentstyle[twocolumn,prl,aps,epsfig]{revtex}

\begin{document}

\wideabs{
\title{Nonlinear Josephson-type oscillations of a driven,
two-component Bose-Einstein condensate}

\author{J. Williams, R. Walser, J. Cooper, E. Cornell$^*$, M. Holland}

\address{JILA, *National Institute for Standards and Technology, and
Department of Physics, University of Colorado, Boulder, CO 80309-0440}

\date{\today}

\maketitle

\begin{abstract}

We propose an experiment that would demonstrate nonlinear
Josephson-type oscillations in the relative population of a driven,
two-component Bose-Einstein condensate. An initial state is prepared
in which two condensates exist in a magnetic trap, each in a different
hyperfine state, where the initial populations and relative phase
between condensates can be controlled within experimental uncertainty.
A weak driving field is then applied, which couples the two internal
states of the atom and consequently transfers atoms back and forth
between condensates. We present a model of this system and investigate
the effect of the mean field on the dynamical evolution.

\end{abstract}
}

An interesting property of a weakly interacting Bose-Einstein
condensate is that it can be ascribed an overall phase that can be
measured relative to another condensate
\cite{phase1,phase2,phase3,phase4,Atec}.  This is a quantum mechanical
effect exhibited on a macroscopic scale. Some recent experiments on
Bose-Einstein condensation (BEC) in dilute alkali vapors have
investigated the relative phase of two overlapping condensates. For
example, in the experiment reported in Ref.~\cite{fringes},
interference fringes in the density of two overlapping condensates
were observed.  More recently, the authors of Ref.~\cite{exp3}
measured the relative phase of two condensates in different hyperfine
states using a technique based on Ramsey's method of separated
oscillating fields \cite{Ramsey}.

A classic experiment that investigates the role of coherence on the
dynamical evolution of two coupled macroscopic quantum systems is the
Josephson-junction experiment, in which a superconducting current of
Cooper-pairs exhibits coherent oscillations \cite{Barone}.  There have
been several proposals suggesting an analogous experiment in the
context of the more recent work being conducted on BEC of dilute
atomic gases 
\cite{Juha,Walls1,Wright,Walls2,Legget1,Legget2,Smerzi1,Smerzi2,Clark,Villain}.
One can imagine preparing two initially isolated condensates in a
double well potential and then lowering the central barrier to allow
coupling between condensates to arise from tunneling.  One could then
observe the time rate-of-change of the relative population, which is
the analogous quantity to the current of Cooper-pairs in the usual
Josephson-junction experiment.  Further interesting effects could then
also be studied, such as the nonlinear effect of the mean field on the
system's behavior \cite{Legget1,Smerzi1}.

We propose a different experiment that would exhibit much the same
physics as in the proposed double-well tunneling experiment but is
based on the work done recently on two-component condensates, where
two different hyperfine states can be populated and confined in the
same trap \cite{exp3,exp1,exp2,Ketterle}. We restrict our attention to
the situation described in Refs.~\cite{exp3} and~\cite{exp2} where the
authors trapped and cooled ${}^{87}$Rb atoms in a magnetic trap below
the critical point for BEC. The trapped atoms were initially in the
$|f=1,m_f=-1 \rangle$ hyperfine state but after condensation the
$|f=2,m_f=1 \rangle$ state could be populated through a two-photon
transition. After this first $\pi/2$-pulse, which lasts a fraction of
the period of the trap, the relative motion of the two condensates
oscillates and damps out to a stationary situation \cite{exp2}. After
applying a second $\pi/2$-pulse, the authors observed a well defined
relative phase that persists even beyond these damped oscillations
\cite{exp3,exp2}.

In light of this observed ``phase rigidity,'' we envision the
following experiment. An initial stationary state is first prepared as
described above by applying a short drive pulse that produces
condensates in both $|1,-1 \rangle$ and $|2,1 \rangle$ states with
known populations. When the transient relative motion has damped out,
the two condensates each sit in different shifted harmonic traps due
to their different magnetic moments, with an overlap region that can
be controlled experimentally. A weak driving field that couples these
two internal states is then applied so that the condensates are
coupled in the overlap region. The time at which this sustained drive
is turned on determines the initial relative phase accumulated by the
condensates, which is measured relative to the accumulated phase of
the driving field \cite{exp3}.  Oscillations in the relative
population will occur, which depend on the initial relative
phase and populations, and will exhibit nonlinear behavior due to the
mean field.

The true system described in Refs.~\cite{exp3} and~\cite{exp2} has
axial symmetry, with the two components separated along the z-axis
axis. We simplify the problem by treating the system in only one
dimension along the z-axis. We make another simplification by treating
temperature zero and neglecting the fluctuations about the mean field
that would give rise to damping.

In order to simplify our notation, we label the hyperfine states as
$|2 \rangle = |2, 1 \rangle$ and $|1 \rangle = |1, -1 \rangle$. In the
presence of a weak external magnetic field these hyperfine states are
separated in frequency by $\omega_0$.  The system is driven by a
two-photon pulse, the strength of which we characterize by the
two-photon Rabi frequency $\Omega$, which we take to be
real-valued. We label the frequency of the two-photon drive by
$\omega_d$, which can be varied to give a finite detuning $\delta =
\omega_d - \omega_0$.  In the following we have made the rotating wave
approximation by dropping the high-frequency terms in the atom-field
interaction.  Finally, we assume both states have a long lifetime
compared to the period of the trap.

After making the above approximations, we carried out standard
mean-field theory on this coupled, two-component system to obtain the
following equations for the time evolution of the condensates in the
rotating frame:
\begin{equation}
\begin{array}{ccc}
i   \left( \! \! \begin{array}{c}  
       \dot{\psi}_2  \\ \dot{\psi}_1  \end{array} \!\! \right)
   \!\!\!  &=& \!\!\! \left( \!\! \begin{array}{cc}  
    H_2^0 + H_2^{\rm{MF}} - 
       \delta/2 & \Omega \\
       \Omega &  H_1^0 + H_1^{\rm{MF}} + \delta/2 
\end{array} \! \right) \!\!   \left( \!\! \begin{array}{c}  
       \psi_2  \\  \psi_1  \end{array} \!\! \right) \,\, .
\end{array}
\label{theone}
\end{equation}
The frequency of each trap is $\omega_z$. We work in the ``natural''
units of the problem, so that time is in units of $1 / \omega_z$,
energy is in terms of the trap level spacing $\hbar \, \omega_z$, and
position is in units of the harmonic oscillator length $z_{\rm{sho}} =
\sqrt{\hbar / m_{\rm{Rb}} \, \omega_z}$. The complex functions
$\psi_i(z,t)$ are the mean field amplitudes of each condensate, where
$i=1,2$. They are normalized to give the populations $N_i(t)$, where
the total number $N =N_2 + N_1$ is constant.

The Hamiltonians appearing in Eq.~(\ref{theone}) describe the free
evolution $H_i^0$ and the mean field interaction $H_i^{\rm{MF}}$ for
each component
\begin{eqnarray}
H_i^0 &=& {1\over{2}} p^2 + {1\over{2}} (z + \gamma_i \,z_0)^2
\nonumber \\ H_i^{\rm{MF}} &=& \lambda_{ii} |\psi_i|^2 + \lambda_{ij}
|\psi_j|^2 \,\,,
\label{Hs}
\end{eqnarray}
where $\gamma_1=1$ and $\gamma_2=-1$, and $z_0$ is the shift of each
trap from the origin. The mean-field strength is characterized by
$\lambda_{ij} = a_{ij}/z_{\rm{sho}}$, which depends on the scattering
length $a_{ij}$ of the collision.  In general there will be three
different values, one for each type of collision in this two-component
gas: $a_{22}$, $a_{11}$, $a_{21}$.

Before presenting results from numerical calculations of
Eq.~(\ref{theone}), it is useful to obtain two different forms of
Eq.~(\ref{theone}) that link this problem to two well known physical
systems in the literature: the standard Rabi problem in quantum optics
\cite{Mandel}, and the Josephson-junction problem in condensed matter
physics \cite{Barone}.

We obtain the equations of motion for the populations $N_i$ and the
coherences $N_{ij} = \int dz \, \psi_i^{*}(z) \psi_j(z)$ from
Eq.~(\ref{theone}) by forming the appropriate products and integrating
over space to yield
\begin{eqnarray}
\dot{N}_2 &=& - i \, \Omega \,\, (N_{21}-N_{12}) \nonumber \\
   \dot{N}_1 &=& + i \, \Omega \,\,(N_{21}-N_{12}) \nonumber \\
   \dot{N}_{21}&=& -i \, \delta \, N_{21} \, + i \, \Lambda(t) \, - \,
   i \, \Omega \, (N_2 - N_1) \,\, ,
\label{number}
\end{eqnarray}
where we define the time-dependent term $\Lambda(t)$ as
\begin{eqnarray}
   \Lambda(t) &=& - 2 \, z_0 \, \int dz \,z \, \psi_2^{*}(z) \psi_1(z)
   \nonumber \\ &+& \int dz \, ( H_2^{\rm{MF}} - H_1^{\rm{MF}}) \,
   \psi_2^{*}(z) \psi_1(z) \,\, .
\label{detuning}
\end{eqnarray}
The equations in Eq.~(\ref{number}) resemble the Bloch equations
describing an undamped, driven two-level atom~\cite{Mandel}.  However,
because the center-of-mass motion is correlated to the internal states
of the atom, the extra term $\Lambda(t)$ appears, which includes the
difference in external potentials between the two states.  

The first term in Eq.~(\ref{detuning}) arises from the difference in
the shifted harmonic traps, which is just linear in $z$. This term
acts as a time-dependent detuning.  As population is transferred from
one condensate to the other, the position of the overlap region
changes due to the mean-field repulsion. This will cause the system to
move in and out of resonance resulting in a suppression of the
transfer of atoms. The second term comes from the difference in
mean-field interactions and would vanish if all three scattering
lengths were exactly degenerate, which can be seen from
\begin{eqnarray}
 H_2^{\rm{MF}} - H_1^{\rm{MF}} &=&
 (\lambda_{22}  -  \lambda_{21})|\psi_2|^2 \nonumber \\
 &-& (\lambda_{11}  -  \lambda_{21})|\psi_1|^2 \,\, .  
\end{eqnarray}

In order to make a link to the standard DC Josephson effect, we must
make some approximations in order to put Eq.~(\ref{theone}) in a
simpler form. For a very weak coupling ($\Omega << 1$) and an initial
state that is the self-consistent solution of the uncoupled system, we
can make the ansatz $\psi_i(z,t) = \sqrt{N_i(t)} \, e^{i \phi_i(t)} \,
\Phi_i(z)$. Here we put the explicit time dependence into the
population $N_i$ and the phase $\phi_i$ of each condensate while
putting the spatial dependence into an adiabatic solution $\Phi_i(z)$
to the undriven system
\begin{equation}
( H_i^0 + H_i^{MF}) \, \Phi_i(z) = \mu_i \, \Phi_i(z) \,\, ,
\label{gp}
\end{equation}
where $\mu_i$ is the chemical potential for each condensate and the
solutions $\Phi_i(z)$ are taken to be real. The chemical potentials
$\mu_i$ and functions $\Phi_i(z)$ vary slowly in time, being
``slaved'' by the populations.

If we substitute this ansatz into Eq.~(\ref{theone}), we obtain the
following equations of motion for the relative population $\eta = (N_2
- N_1)/N$ and the relative phase $\phi = (\phi_2 - \phi_1)$
\begin{eqnarray}
  \dot{\eta} &=& -k \, (1-\eta^2)^{1/2} \sin\phi \nonumber \\
  \dot{\phi} &=& -[(\mu_2 - \mu_1) - \delta]
             + k \, \eta \, (1-\eta^2)^{-1/2} \cos\phi \,\, ,
\label{jj}
\end{eqnarray}
where $k = 2\,\Omega \int {\rm{dz}}\, \Phi_2(z) \Phi_1(z)$ is
proportional to the overlap of the condensates and so also varies
slowly in time.  These are non-linear versions of the usual
Josephson-junction equations \cite{Barone} and are nearly identical in
form to those obtained in Refs.~\cite{Smerzi1} and~\cite{Smerzi2}
describing the double-well tunneling problem.  The major difference is
that in the double-well trap, the condensates are well separated,
allowing the authors in Refs.~\cite{Smerzi1} and~\cite{Smerzi2} to
neglect the mean field in the interaction region of the barrier.  In
contrast, the interaction between condensates due to their significant
overlap plays an important role in the evolution of the system
described in this letter. In particular, it is this mutual interaction
that causes the system to move out of resonance.

We now show some results of calculations of both the exact solution
given by Eq.~(\ref{theone}), and the approximate solution given by
Eq.~(\ref{jj}). The main purpose of the present calculations is to use
realistic parameters to investigate the effect the mean field has on
the evolution of the system. These parameters are listed in Table 1.
\begin{figure}
  \centerline{\epsfig{file=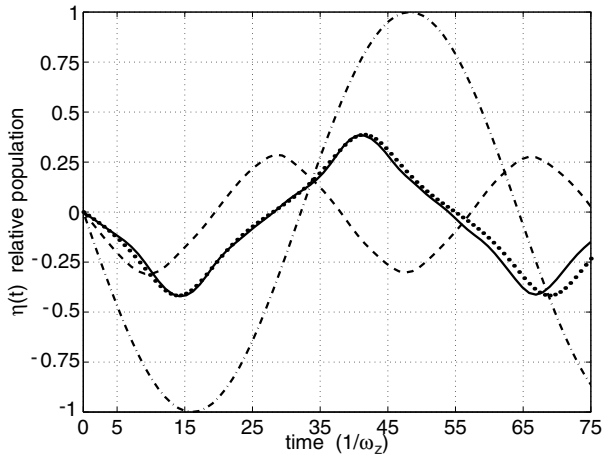,height=2.45in}}
\caption{This plot shows oscillations of the relative population. The
dashed-dot line corresponds to the case where the mean-field
interaction has been set to zero, the initial relative phase is
$\phi(0) = \pi/2$, and $\delta = 0$. The dashed line is for the same
initial relative phase of $\pi/2$, but with the mean-field interaction
turned on and $\delta = -0.39$. The solid line is for
$\phi(0)=3\pi/16$ and $\delta = -0.39$. The three lines described were
solutions of Eq.~(\ref{theone}) whereas the dotted line is a solution
of Eq.~(\ref{jj}) with the same parameters as in the solid line.}
\end{figure}

In Fig. 1 we show four curves that are described in the caption.  As a
point of reference, we plot the solution of Eq.~(\ref{theone}) with
the mean-field terms set to zero. This is the standard Rabi solution,
but here the Rabi frequency is given by $\omega_{\rm{R}}=2 \, \Omega
\, \int dz \, \Phi_2(z) \Phi_1(z)$, which includes the
Frank-Condon-type overlap of the condensate wavefunctions.  However,
when we turn on the mean-field interaction using the parameters given
in Table 1, and set $\delta = -0.39$ so that the system is initially
begin driven resonantly, the amplitude is suppressed and the frequency
has increased (the dashed line). As $\phi(0)$ is decreased, the
amplitude increases, as illustrated by the solid curve where $\phi(0)
= 3 \pi /16$.  Also, as $\phi(0)$ is decreased, the presence of higher
harmonics becomes stronger, as one can see in the shape of the solid
line.

We also plot the adiabatic solution given by Eq.~(\ref{jj}) for the
case $\phi(0) = 3 \pi /16$ (the dotted line), where Eq.~(\ref{gp}) is
solved self-consistently in each time step.  In this case the
adiabatic solution agrees quite well with the exact solution (solid
line) given by Eq.~(\ref{theone}). The validity of the two-level,
adiabatic solution depends on the structure of the evolving spectrum
of this nonlinear system. In particular, one must compare the
time-rate-of-change of the Hamiltonian to the spacing between the
instantaneous eigenmodes of the dressed basis \cite{Messiah}.  These
quantities will depend on the size of the mean-field interaction, the
strength of the coupling, given by $2\,\Omega \, \int dz \, \Psi_2(z)
\Psi_1(z)$, and also on the detuning $\delta$.  It should also be
noted that in the true system, the confining potential along the
xy-plane is weaker than that along the z-axis, which should imply a
more stringent criterion for adiabaticity than in the one-dimensional
case considered.

\begin{figure}
  \centerline{\epsfig{file=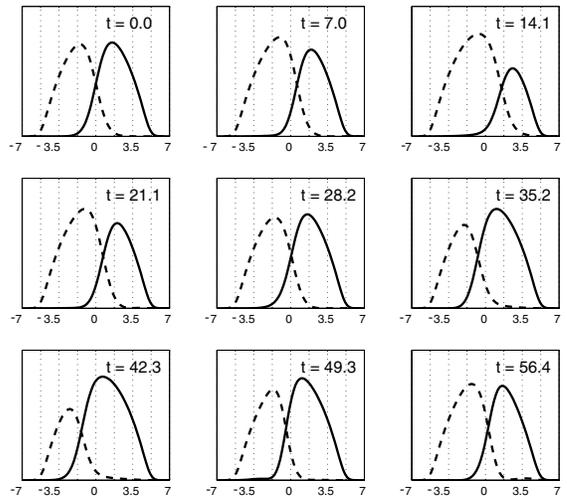,height=2.65in}}
\vspace{.08in}
\caption{This plot shows the time-evolution of the density of each
component. The x-axis is the position $z$, and the time $t$ of each
snapshot is shown. This case corresponds to the solid line plotted in
Fig. 1. The detuning is $\delta = -0.39$, chosen so that the system is
initially being driven on resonance.}
\end{figure}
We plot a snapshot evolution of the densities of the condensates in
Fig. 2 in order to show that the effect of the mean field is to push
the system out of resonance as population is transferred between
condensates.  The case considered in Fig. 2 corresponds to the solid
line in Fig. 1.  The detuning was chosen so as to compensate for the
initial value of the term $\Lambda(0)$ in Eq.~(\ref{number}), which
represents the difference in external potentials, so that initially
the system is being driven on resonance.  However, as the system
evolves, the first term in Eq.~(\ref{detuning}) gets larger since the
$|1\rangle$ condensate is pushing the $|2\rangle$ condensate away from
the center of the trap.  This causes the region of overlap to be
displaced from the origin so that the system is no longer being driven
on resonance.  At $t=14.1$ the region of overlap is centered at about
$z=1.5$, at which time $| \Lambda(t)/N_{21}(t)|=0.63$, compared to the
initial value $| \Lambda(0)/N_{21}(0)|=0.39$. This reduces the
effectiveness of the drive and accounts for the suppression of the
amplitude of oscillation in the relative population plotted in Fig. 1.

\begin{figure}
  \centerline{\epsfig{file=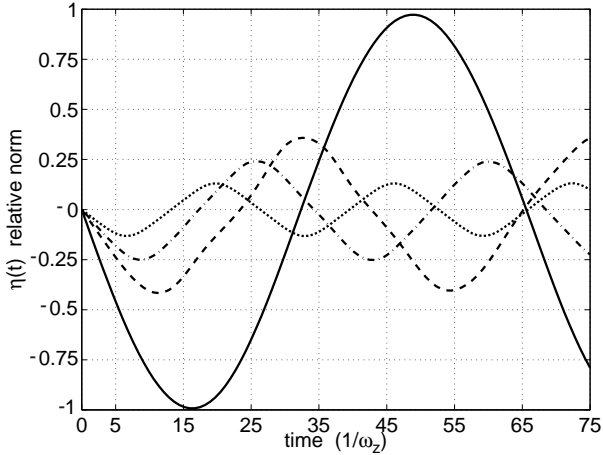,height=2.45in}}
\caption{This plot shows that the effect of the displacement $z_0$ on
the system is to suppress the amplitude of oscillation.  The four
curves represent increasing values of $z_0$: 0 (solid), 0.1 (dashed),
0.2 (dash-dotted), and 0.5 (dotted). In each case, the detuning
$\delta$ is chosen so that the system is initially being driven
resonantly, and the initial phase is $\phi(0) = \pi/2$.}
\label{fig1}
\end{figure} 
In Fig. 3 we vary the displacement $z_0$ for four different values.
The amplitude and period of oscillation both decrease as the
displacement between the traps is made larger, while the overall shape
does not vary much.  In other words, the effective detuning caused by
$\Lambda(t)$ becomes more pronounced for larger trap
displacements. There is another effect of increasing $z_0$: the region
of overlap between the condensates decreases as the traps are pulled
apart, which weakens the coupling.  One might expect the period of
oscillation to increase in this case, however, the plot displays a
decreasing period as $z_0$ is increased. This indicates that the
time-dependent detuning due to $\Lambda(t)$ has a stronger effect.

In this letter we have described a physical system based on the
experiments reported in Refs.~\cite{exp1,exp2,exp3} that would exhibit
non-linear Josephson-like oscillations in the relative population
between a driven two-component condensate. From our calculations we
have observed the effect of the mean field, which acts to suppress the
transfer of atoms between overlapping condensates in separated
harmonic traps and gives rise to nonlinear oscillations in the
relative population.  We have also found that, for equal populations
initially $\eta(0)=0$, the spectrum of Fourier components comprising
the oscillation depends on the initial relative phase $\phi(0)$.

We would like to thank Marilu` Chiofalo for helpful discussions. We
also thank the National Science Foundation for funding this work.
\bibliographystyle{prsty}
\bibliography{Paperbib}

\begin{thebibliography}{10}

\bibitem{phase1}
J. Javanainen and S. Yoo, \prl {\bf 76},  161  (1996).

\bibitem{phase2}
M. Naraschewski {\it et~al.}, \pra {\bf 54},  2185  (1996).

\bibitem{phase3}
S. Barnett, K. Burnett, and J. Vaccaro, Journal of Research of NIST {\bf 101},
  593  (1996).

\bibitem{phase4}
Y. Castin and J. Dalibard, \pra {\bf 55},  4330  (1997).

\bibitem{Atec}
A. Imamoglu and T. Kennedy, \pra {\bf 55},  R849  (1997).

\bibitem{fringes}
M.~R. Andrews {\it et~al.}, Science {\bf 275},  637  (1997).

\bibitem{exp3}
D. Hall, M. Matthews, C. Wieman, and E. Cornell, \prl {\bf 81},  1543  (1998).

\bibitem{Ramsey}
N. Ramsey, {\em Molecular Beams} (Clarendon Press, Oxford, 1956).

\bibitem{Barone}
A. Barone and G. Paterno, {\em Physics and Applications of the Josephson
  Effect} (Wiley, New York, 1982).

\bibitem{Juha}
J. Javanainen, \prl {\bf 57},  3164  (1986).

\bibitem{Walls1}
M. Jack, M. Collett, and D. Walls, \pra {\bf 54},  R4625  (1996).

\bibitem{Wright}
G. Milburn, J. Corney, E. Wright, and D. Walls, \pra {\bf 55},  4318  (1997).

\bibitem{Walls2}
J. Ruostekoski and D. Walls, \pra {\bf 56},  2996  (1997).

\bibitem{Legget1}
I. Zapata, F. Sols, and A. Leggett, \pra {\bf 57},  R28  (1998).

\bibitem{Legget2}
A.~J. Leggett, Journal of Low Temperature Physics {\bf 110},  719  (1998).

\bibitem{Smerzi1}
A. Smerzi, S. Fantoni, S. Giovanazzi, and S. Shenoy, \prl {\bf 79},  4950
  (1997).

\bibitem{Smerzi2}
S. Raghavan, A. Smerzi, S. Fantoni, and S.~R. Shenoy, cond-mat/9706220  .

\bibitem{Clark}
W. Reinhardt and C. Clark, J. Phys. B:At. Mol. Opt. Phys {\bf 30},  L785
  (1997).

\bibitem{Villain}
P. Villain and M. Lewenstein, quant-ph/9808017  .

\bibitem{exp1}
M. Matthews {\it et~al.}, \prl {\bf 81},  243  (1998).

\bibitem{exp2}
D. Hall {\it et~al.}, \prl {\bf 81},  1539  (1998).

\bibitem{Ketterle}
D. Stamper-Kurn {\it et~al.}, \prl {\bf 80},  2027  (1998).

\bibitem{Mandel}
L. Mandel and E. Wolf, {\em Optical Coherence and Quantum Optics} (Cambridge,
  New York, 1995).

\bibitem{Messiah}
A. Messiah, {\em Quantum Mechanics} (John Wiley and Sons, New York, 1966),
  Vol.~2.

\end{thebibliography}

\begin{table*}[Top]
\begin{center}
\caption{This is a table showing the values used for the various
physical parameters appearing in our calculations. The scattering
lengths are taken from Ref.~[21]. Since we are treating the system in
1D, $N$ is not the actual population, but was chosen to produce a
realistic mean-field interaction for $5 \times 10^5$ atoms [7].}
\begin{tabular}{llcll}
  $N$ & $2.3 \times 10^4$ & \vline & $\nu_z$ & $60 \, {\rm{Hz}}$ \\
$a_{21}$ & $5.5(3) \,{\rm{nm}}$ & \vline & $\Omega/2\pi$ & $1/20 \,\,
\nu_z$ \\ $a_{22}$ & $0.97 \, a_{21}$ & \vline & $z_{\rm{sho}}$ & $1.4
\, \mu$m \\ $a_{11}$ & $1.03 \, a_{21}$ & \vline & $z_0$ & $0.15 \,
z_{\rm{sho}}$ \\ \end{tabular} \end{center} \label{table1}
\end{table*} \end{document}